\documentclass[a4paper]{report}
\usepackage[utf8]{inputenc}
\usepackage[T1]{fontenc}
\usepackage{RJournal}
\usepackage{amsmath,amssymb,array}
\usepackage{booktabs}

\providecommand{\tightlist}{%
  \setlength{\itemsep}{0pt}\setlength{\parskip}{0pt}}

\graphicspath{{./figures/}}

\newcommand{\orcid}[1]{\href{https://orcid.org/#1}{\includegraphics[height=\fontcharht\font`\0]{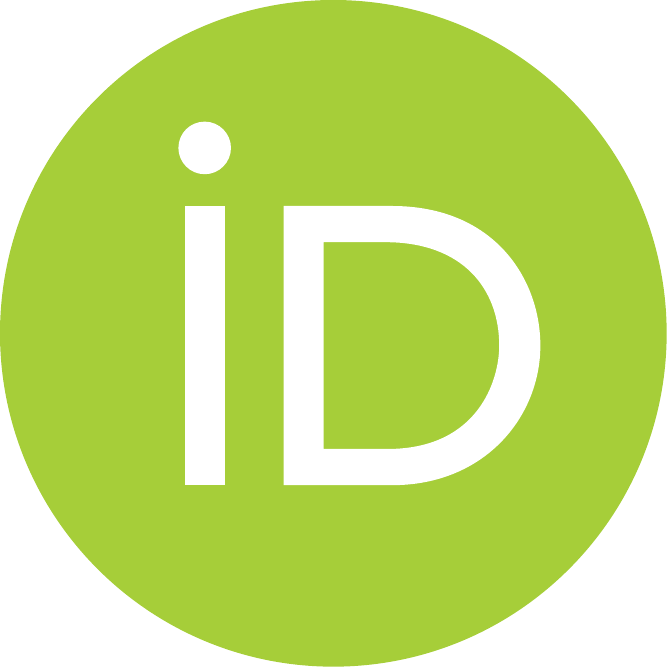} #1}}

\begin{document}

%% do not edit, for illustration only
\sectionhead{Preprint}
\volume{}
\volnumber{}
\year{}
\month{}

\begin{article}
  % !TeX root = RJwrapper.tex
\title{The Rockerverse: Packages and Applications for Containerisation
with R}
\author{by Daniel Nüst, Dirk Eddelbuettel, Dom Bennett, Robrecht
Cannoodt, Dav Clark, Gergely Daróczi, Mark Edmondson, Colin Fay, Ellis
Hughes, Lars Kjeldgaard, Sean Lopp, Ben Marwick, Heather
Nolis, Jacqueline Nolis, Hong Ooi, Karthik Ram, Noam Ross, Lori
Shepherd, Péter Sólymos, Tyson Lee Swetnam, Nitesh Turaga, Charlotte Van
Petegem, Jason Williams, Craig Willis, Nan Xiao}

\maketitle

\abstract{%
The Rocker Project provides widely used Docker images for R across
different application scenarios. This article surveys downstream
projects that build upon the Rocker Project images and presents the
current state of R packages for managing Docker images and controlling
containers. These use cases cover diverse topics such as package
development, reproducible research, collaborative work, cloud-based data
processing, and production deployment of services. The variety of
applications demonstrates the power of the Rocker Project specifically
and containerisation in general. Across the diverse ways to use
containers, we identified common themes: reproducible environments,
scalability and efficiency, and portability across clouds. We conclude
that the current growth and diversification of use cases is likely to
continue its positive impact, but see the need for consolidating the
Rockerverse ecosystem of packages, developing common practices for
applications, and exploring alternative containerisation software.
}

\hypertarget{introduction}{%
\section{Introduction}\label{introduction}}

\label{intro}

The R community continues to grow. This can be seen in the number of new
packages on CRAN, which is still on growing exponentially
\citep{cran:2019}, but also in the numbers of conferences, open
educational resources, meetups, unconferences, and companies that are
adopting R, as exemplified by the useR! conference
series\footnote{\href{https://www.r-project.org/conferences/}{https://www.r-project.org/conferences/}},
the global growth of the R and R-Ladies user
groups\footnote{\href{https://www.r-consortium.org/blog/2019/09/09/r-community-explorer-r-user-groups}{https://www.r-consortium.org/blog/2019/09/09/r-community-explorer-r-user-groups}, \href{https://www.r-consortium.org/blog/2019/08/12/r-community-explorer}{https://www.r-consortium.org/blog/2019/08/12/r-community-explorer}},
or the foundation and impact of the
R~Consortium\footnote{\href{https://www.r-consortium.org/news/announcements}{https://www.r-consortium.org/news/announcements}, \href{https://www.r-consortium.org/blog/2019/11/14/data-driven-tracking-and-discovery-of-r-consortium-activities}{https://www.r-consortium.org/blog/2019/11/14/data-driven-tracking-and-discovery-of-r-consortium-activities}}.
These trends cement the role of R as the \emph{lingua~franca} of
statistics, data visualisation, and computational research. The last few
years, coinciding with the rise of R, have also seen the rise of
\href{https://en.wikipedia.org/wiki/Docker_(software)}{Docker} as a
general tool for distributing and deploying of server applications---in
fact, Docker can be called the \emph{lingua~franca} of describing
computing environments and packaging software. Combining both these
topics, the \emph{Rocker~Project}
(\url{https://www.rocker-project.org/}) provides Docker images with R
(see the next section for more details). The considerable uptake and
continued evolution of the Rocker~Project has led to numerous projects
that extend or build upon Rocker images, ranging from
reproducible\footnote{"Reproducible" in the sense of the \emph{Claerbout/Donoho/Peng} terminology \citep{barba_terminologies_2018}.}
research to production deployments. As such, this article presents what
we may call the \emph{Rockerverse} of projects across all development
stages: early demonstrations, working prototypes, and mature products.
We also introduce related activities that connect the R language and
environment with other containerisation solutions. Our main contribution
is a coherent picture of the current status of using containers in,
with, and for R.

The article continues with a brief introduction of containerisation
basics and the Rocker~Project, followed by use cases and applications,
starting with the R packages specifically for interacting with Docker,
next the second-level packages that use containers indirectly or only
for specific features, and finally some complex use cases that leverage
containers. We conclude by reflecting on the landscape of packages and
applications and point out future directions of development.

\hypertarget{containerisation-and-rocker}{%
\section{Containerisation and
Rocker}\label{containerisation-and-rocker}}

\label{containerisation} \label{rocker}

Docker, an application and service provided by the eponymous company,
has, in just a few short years, risen to prominence for developing,
testing, deploying and distributing computer software
\citep[cf.][]{datadog_8_2018,munoz_history_2019}. While related
approaches exist, such as
LXC\footnote{\href{https://en.wikipedia.org/wiki/LXC}{https://en.wikipedia.org/wiki/LXC}}
or Singularity \citep{kurtzer_singularity_2017}, Docker has become
synonymous with ``containerisation''---the method of taking software
artefacts and bundling them in such a way that use becomes standardized
and portable across operating systems. In doing so, Docker had
recognised and validated the importance of one very important thread
that had been emerging, namely virtualisation. By allowing (one or
possibly) multiple applications or services to run concurrently on one
host machine without any fear of interference between them, Docker
provides an important scalability opportunity. Beyond this though,
Docker has improved this compartmentalisation by accessing the host
system---generally Linux---through a much thinner and smaller shim than
a full operating system emulation or virtualisation. This
containerisation, also called operating-system-level virtualisation
\citep{wikipedia_contributors_os-level_2020}, makes efficient use of
operating system resources \citep{felter_updated_2015} and allows
another order of magnitude in terms of scalability of deployment
\citep[cf.][]{datadog_8_2018}, because virtualisation may emulate a
whole operating system, a container typically runs only one process. The
single process together with sharing the host's kernel results in a
reduced footprint and faster start times. While Docker makes use of
Linux kernel features, it has become important enough that some required
aspects of running Docker have been added to other operating systems so
that those systems can more efficiently support Docker
\citep{microsoft_linux_2019}. The success of Docker has even paved the
way for industry collaboration and standardisation
\citep{oci_open_2019}.

The key accomplishment of Docker as an ``application'' is to make a
``bundled'' aggregation of software, the so-called ``image'', available
to any system equipped to run Docker, without requiring much else from
the host besides the actual Docker application installation. This is a
rather attractive proposition, and Docker's very easy to operate user
interface has led to widespread adoption and use of Docker in a variety
of domains, e.g., cloud computing infrastructure
\citep[e.g.,][]{Bernstein2014}, data science
\citep[e.g.,][]{boettiger_introduction_2015}, and edge computing
\citep[e.g.,][]{alam_orchestration_2018}. It has also proven to be a
natural match for ``cloud deployment'' which runs, or at least appears
to run, ``seamlessly'' without much explicit reference to the underlying
machine, architecture or operating system: Containers are portable and
can be deployed with very little dependencies on the host system---only
the container runtime is required. These Docker images are normally
built from plain text documents called \texttt{Dockerfile}s; a
\texttt{Dockerfile} has a specific set of instructions to create and
document a well-defined environment, i.e., install specific software and
expose specific ports.

For statistical computing and analysis centred around R, the
\textbf{Rocker~Project} has provided a variety of Docker containers
since it began in 2014 \citep{RJ-2017-065}. The Rocker~Project provides
several lines of containers spanning from building blocks with
\texttt{R-release} or \texttt{R-devel}, via containers with
\href{https://rstudio.com/products/rstudio/}{RStudio~Server} and
\href{https://rstudio.com/products/shiny/shiny-server/}{Shiny~Server},
to domain-specific containers such as
\href{https://github.com/rocker-org/geospatial}{\texttt{rocker/geospatial}}
\citep{rocker_geospatial_2019}. These containers form \emph{image
stacks}, building on top of each other for easier maintainability (i.e.,
smaller \texttt{Dockerfile}s), better composability, and to reduce build
time. Also of note is a series of ``versioned'' containers which match
the R release they contain with the \emph{then-current} set of packages
via the MRAN Snapshot views of CRAN \citep{microsoft_cran_2019}. The
Rocker~Project's impact and importance was acknowledged by the Chan
Zuckerberg Initiative's \emph{Essential Open Source Software for
Science}, which provides funding for the project's sustainable
maintenance, community growth, and targeting new hardware platforms
including GPUs \citep{chan_zuckerberg_initiative_maintaining_2019}.

Docker is not the only containerisation software. \textbf{Singularity}
stems from the domain of high-performance computing
\citep{kurtzer_singularity_2017} and can also run Docker images. Rocker
images work out of the box if the main process is R, e.g., in
\texttt{rocker/r-base}, but Singularity does not succeed in running
images where there is an init script, e.g., in containers that by
default run RStudio~Server. In the latter case, a \texttt{Singularity}
file, a recipe akin to a \texttt{Dockerfile}, needs to be used to make
necessary adjustments. To date, no comparable image stack to the Rocker
Project's images exists on
\href{https://singularity-hub.org/}{Singularity Hub}. A further tool for
running containers is
\href{https://github.com/containers/libpod}{\textbf{podman}}, which can
also build \texttt{Dockerfile}s and run Docker images. Proof of concepts
exists for using podman to build and run Rocker
containers\footnote{See \href{https://github.com/nuest/rodman}{https://github.com/nuest/rodman} and \href{https://github.com/rocker-org/rocker-versioned/issues/187}{https://github.com/rocker-org/rocker-versioned/issues/187}},
but the prevalence of Docker, especially in the broader user community
beyond experts or niche systems and the vast amount of blog posts and
courses for Docker currently cap specific development efforts for both
Singularity and podman in the R community. This might quickly change if
the usability and spread of Singularity or podman increase, or if
security features such as rootless/unprivileged containers, which both
these tools support out of the box, become more sought after.

\hypertarget{interfaces-for-docker-in-r}{%
\section{Interfaces for Docker in R}\label{interfaces-for-docker-in-r}}

\label{interfaces}

Users interact with the Docker daemon typically through the
\href{https://docs.docker.com/engine/reference/commandline/cli/}{Docker
Command Line Interface} (Docker CLI). However, moving back and forth
between an R console and the command line can create friction in
workflows and reduce reproducibility because of manual steps. A number
of first-order R packages provide an interface to the Docker CLI,
allowing for the interaction with the Docker CLI from an R console.
Table~\ref{tab:clients} gives an overview of packages with client
functionality, each of which provides functions for interacting with the
Docker daemon. The packages focus on different aspects and support
different stages of a container's life cycle. As such, the choice of
which package is most useful depends on the use case at hand as well as
on the user's level of expertise.

\begin{table}
\centering
\begin{tabular}{llllllll}
  \toprule
\rotatebox{-90}{Functionality} & \rotatebox{-90}{AzureContainers} & \rotatebox{-90}{babelwhale} & \rotatebox{-90}{dockermachine} & \rotatebox{-90}{dockyard} & \rotatebox{-90}{googleCloudRunner} & \rotatebox{-90}{harbor} & \rotatebox{-90}{stevedore }\\ 
  \midrule
Generate a Dockerfile &  &  &  & \checkmark &  &  &  \\ 
   \midrule
Build an image & \checkmark &  &  & \checkmark & \checkmark &  &  \\ 
   \midrule
Execute a container locally or remotely & \checkmark & \checkmark & \checkmark & \checkmark & \checkmark & \checkmark & \checkmark \\ 
   \midrule
Deploy or manage instances in the cloud & \checkmark &  & \checkmark &  & \checkmark & \checkmark & \checkmark \\ 
   \midrule
Interact with an instance (e.g., file transfer) &  & \checkmark & \checkmark &  &  &  & \checkmark \\ 
   \midrule
Manage storage of images &  &  &  &  &  & \checkmark & \checkmark \\ 
   \midrule
Supports Docker and Singularity &  & \checkmark &  &  &  &  &  \\ 
   \midrule
Direct access to Docker API instead of using the CLI &  &  &  &  &  &  & \checkmark \\ 
   \midrule
Installing Docker software &  &  & \checkmark &  &  &  &  \\ 
   \bottomrule
\end{tabular}
\caption{R packages with Docker client functionality.} 
\label{tab:clients}
\end{table}

\textbf{\pkg{harbor}} (\url{https://github.com/wch/harbor}) is no longer
actively maintained, but it should be honourably mentioned as the first
R package for managing Docker images and containers. It uses the
\CRANpkg{sys} package \citep{cran_sys} to run system commands against
the Docker CLI, both locally and through an SSH connection, and it has
convenience functions, e.g., for listing and removing containers/images
and for accessing logs. The outputs of container executions are
converted to appropriate R types. The Docker CLI's basic functionality,
although it evolves quickly and with little concern for avoiding
breaking changes, has remained unchanged in core functions, meaning that
a core function such as
\code{harbor::docker\_run(image = "hello-world")} still works despite
its stopped development.

\textbf{\CRANpkg{stevedore}} is currently the most powerful Docker
client in R \citep{cran_stevedore}. It interfaces with the Docker daemon
over the Docker HTTP
API\footnote{\href{https://docs.docker.com/engine/api/latest/}{https://docs.docker.com/engine/api/latest/}}
via a Unix socket on Linux or MacOS, over a named pipe on Windows, or
over an HTTP/TCP connection. The package is the only one not using
system calls to the \texttt{docker} CLI tool for managing images and
containers. The package thereby enables connections to remote Docker
instances without direct configuration of the local Docker daemon.
Furthermore using the API gives access to information in a structured
way, is system independent, and is likely more reliable than parsing
command line output. \pkg{stevedore}'s own interface is automatically
generated based on the OpenAPI specification of the Docker daemon, but
it is still similar to the Docker CLI. The interface is similar to R6
objects, in that an object of class \code{"stevedore\_object"} has a
number of functions attached to it that can be called, and multiple
specific versions of the Docker API can be supported thanks to the
automatic
generation\footnote{See \href{https://github.com/richfitz/stevedore/blob/master/development.md}{https://github.com/richfitz/stevedore/blob/master/development.md}.}.

\textbf{\CRANpkg{AzureContainers}} is an interface to a number of
container-related services in Microsoft's
\href{https://azure.microsoft.com/}{Azure Cloud}
\citep{AzureContainers_2019}. While it is mainly intended for working
with Azure, as a convenience feature it includes lightweight,
cross-platform shells to Docker and Kubernetes (tools \texttt{kubectl}
and \texttt{helm}). These can be used to create and manage arbitrary
Docker images and containers, as well as Kubernetes clusters on any
platform or cloud service.

\textbf{\CRANpkg{googleCloudRunner}} is an interface with
\href{https://cloud.google.com/}{Google Cloud Platform}
container-related services, with tools to make it easier for R users to
interact with them for common use cases \citep{cran:googleCloudRunner}.
It includes deployment functions for creating R APIs using the
Docker-based \href{https://cloud.run}{Cloud Run} service. Users can
create long running batch jobs calling any Docker image including Rocker
via \href{https://cloud.google.com/cloud-build/}{Cloud Build} and
schedule services using \href{https://cloud.google.com/scheduler/}{Cloud
Scheduler}.

\textbf{\CRANpkg{babelwhale}} provides a unified interface to interact
with Docker and Singularity containers \citep{cannoodt_babelwhale_2019}.
Users can, for example, execute a command inside a container, mount a
volume, or copy a file with the same R commands for both container
runtimes.

\textbf{\pkg{dockyard}}
(\url{https://github.com/thebioengineer/dockyard}) has the goal of
lowering the barrier to creating \texttt{Dockerfile}s, building Docker
images, and deploying Docker containers. The package follows the
increasingly used piping paradigm of the Tidyverse-style
\citep{wickham_welcome_2019} of programming for chaining R functions
representing the instructions in a \texttt{Dockerfile}. An existing
\texttt{Dockerfile} can be used as a template. \pkg{dockyard} also
includes wrappers for common steps, such as installing an R package or
copying files, and provides built-in functions for building an image and
running a container, which make Docker more approachable within a single
R-based user interface.

\textbf{\pkg{dockermachine}}
(\url{https://github.com/cboettig/dockermachine}) is an R package to
provide a convenient interface to
\href{https://docs.docker.com/machine/overview/}{Docker~Machine} from R.
The CLI tool \texttt{docker-machine} allows users to create and manage a
virtual host on local computers, local data centres, or at cloud
providers. A local Docker installation can be configured to
transparently forward all commands issued on the local Docker CLI to a
selected (remote) virtual host. Docker~Machine was especially crucial
for local use in the early days of Docker, when no native support was
available for Mac or Windows computers, but it remains relevant for
provisioning on remote systems. The package has not received any updates
for two years, but it is functional with a current version of
\texttt{docker-machine} (\texttt{0.16.2}). It potentially lowers the
barriers for R users to run containers on various hosts if they perceive
that using the Docker~Machine CLI directly as a barrier and it enables
scripted workflows with remote processing.

\hypertarget{use-cases-and-applications}{%
\section{Use cases and applications}\label{use-cases-and-applications}}

\label{applications}

\hypertarget{image-stacks-for-communities-of-practice}{%
\subsection{Image stacks for communities of
practice}\label{image-stacks-for-communities-of-practice}}

\textbf{Bioconductor} (\url{https://bioconductor.org/}) is an
open-source, open development project for the analysis and comprehension
of genomic data \citep{gentleman_bioconductor_2004}. As of October 30th
2019, the project consists of 1823 R software packages, as well as
packages containing annotation or experiment data. \emph{Bioconductor}
has a semi-annual release cycle, where each release is associated with a
particular version of R, and Docker images are provided for current and
past versions of \emph{Bioconductor} for convenience and
reproducibility. All images, which are described on the
\emph{Bioconductor} web site (see
\url{https://bioconductor.org/help/docker/}), are created with
\texttt{Dockerfile}s maintained on GitHub and distributed through
Docker~Hub\footnote{See \href{https://github.com/Bioconductor/bioconductor_docker}{https://github.com/Bioconductor/bioconductor\_docker} and \href{https://hub.docker.com/u/bioconductor}{https://hub.docker.com/u/bioconductor} respectively.}.
\emph{Bioconductor}'s ``base'' Docker images are built on top of the
\texttt{rocker/rstudio} image. \emph{Bioconductor} installs packages
based on the R version in combination with the Bioconductor version and,
therefore, uses Bioconductor version tagging \texttt{devel} and
\texttt{RELEASE\_X\_Y}, e.g., \texttt{RELEASE\_3\_10}. Past and current
combinations of R and \emph{Bioconductor} will therefore be accessible
via specific image tags.

The \emph{Bioconductor} \texttt{Dockerfile} selects the desired R
version from Rocker images, adds required system dependencies, and uses
the \CRANpkg{BiocManager} package for installing appropriate versions of
\emph{Bioconductor} packages \citep{cran_biocmanager}. A strength of
this approach is that the responsibility for complex software
configuration and customization is shifted from the user to the
experienced \emph{Bioconductor} core team. However, a recent audit of
the \emph{Bioconductor} image stack \texttt{Dockerfile} led to the
deprecation of several community-maintained images, because the numerous
specific images became too hard to understand, complex to maintain, and
cumbersome to customise. As part of the simplification, a recent
innovation is the \texttt{bioconductor\_docker:devel} image, which
emulates the \emph{Bioconductor} environment for nightly builds as
closely as possible. This image contains the environment variables and
the system dependencies needed to install and check almost all
\emph{Bioconductor} software packages (1813 out of 1823). It saves users
and package developers from creating this environment themselves.
Furthermore, the image is configured so that \code{.libPaths()} has
\samp{/usr/local/lib/R/host-site-library} as the first location. Users
mounting a location on the host file system to this location can
persistently manage installed packages across Docker containers or image
updates. Many R users pursue flexible workflows tailored to particular
analysis needs rather than standardized workflows. The new
\texttt{bioconductor\_docker} image is well suited for this preference,
while \texttt{bioconductor\_docker:devel} provides developers with a
test environment close to \emph{Bioconductor}'s build system.

\label{datascience} \textbf{Data science} is a widely discussed topic in
all academic disciplines \citep[e.g.,][]{donoho_50_2017}. These
discussions have shed light on the tools and craftspersonship behind the
analysis of data with computational methods. The practice of data
science often involves combining tools and software stacks and requires
a cross-cutting skillset. This complexity and an inherent concern for
openness and reproducibility in the data science community has led to
Docker being used widely. The remainder of this section presents example
Docker images and image stacks featuring R intended for data science.

\begin{itemize}
\tightlist
\item
  The \href{https://github.com/jupyter/docker-stacks/}{\emph{Jupyter
  Docker Stacks}} project is a set of ready-to-run Docker images
  containing Jupyter applications and interactive computing tools
  \citep{project_jupyter_jupyter_2018}. The \texttt{jupyter/r-notebook}
  image includes R and ``popular packages'', and naturally also the
  IRKernel (\url{https://irkernel.github.io/}), an R kernel for Jupyter,
  so that Jupyter Notebooks can contain R code cells. R is also included
  in the catchall \texttt{jupyter/datascience-notebook}
  image\footnote{\href{https://jupyter-docker-stacks.readthedocs.io/en/latest/using/selecting.html}{https://jupyter-docker-stacks.readthedocs.io/en/latest/using/selecting.html}}.
  For example, these images allow users to quickly start a Jupyter
  Notebook server locally or build their own specialised images on top
  of stable toolsets. R is installed using the Conda package
  manager\footnote{\href{https://conda.io/}{https://conda.io/}}, which
  can manage environments for various programming languages, pinning
  both the R version and the versions of R
  packages\footnote{See \texttt{jupyter/datascience-notebook}'s \texttt{Dockerfile} at \href{https://github.com/jupyter/docker-stacks/blob/master/datascience-notebook/Dockerfile\#L47}{https://github.com/jupyter/docker-stacks/blob/master/datascience-notebook/Dockerfile\#L47}.}.
\item
  \emph{Kaggle} provides the
  \href{https://hub.docker.com/r/kaggle/rstats}{\texttt{gcr.io/kaggle-images/rstats}}
  image (previously \texttt{kaggle/rstats}) and
  \href{https://github.com/Kaggle/docker-rstats}{corresponding
  \texttt{Dockerfile}} for usage in their Machine Learning competitions
  and easy access to the associated datasets. It includes machine
  learning libraries such as Tensorflow and Keras (see also image
  \texttt{rocker/ml} in Section~\nameref{rocker-gpu}), and it also
  configures the \CRANpkg{reticulate} package \citep{cran_reticulate}.
  The image uses a base image with \emph{all packages from CRAN},
  \texttt{gcr.io/kaggle-images/rcran}, which requires a Google Cloud
  Build because Docker~Hub would time
  out\footnote{Originally, a stacked collection of over 20~images with automated builds on Docker~Hub was used, see \href{https://web.archive.org/web/20190606043353/http://blog.kaggle.com/2016/02/05/how-to-get-started-with-data-science-in-containers/}{https://web.archive.org/web/20190606043353/http://blog.kaggle.com/2016/02/05/how-to-get-started-with-data-science-in-containers/} and \href{https://hub.docker.com/r/kaggle/rcran/dockerfile}{https://hub.docker.com/r/kaggle/rcran/dockerfile}}.
  The final extracted image size is over 25~GB, which calls into
  question whether having everything available is actually convenient.
\item
  The \href{https://radiant-rstats.github.io/docs/}{\emph{Radiant}}
  project provides several images, e.g.,
  \href{https://hub.docker.com/r/vnijs/rsm-msba-spark}{\texttt{vnijs/rsm-msba-spark}},
  for their browser-based business analytics interface based on
  \CRANpkg{Shiny} \citep{cran_shiny}, and for use in education as part
  of an MSc
  course\footnote{`Dockerfile` available on GitHub: \href{https://github.com/radiant-rstats/docker}{https://github.com/radiant-rstats/docker}.}.
  As data science often applies a multitude of tools, this image favours
  inclusion over selection and features Python, Postgres, JupyterLab and
  Visual Studio Code besides R and RStudio, bringing the image size up
  to 9~GB.
\item
  \emph{Gigantum} (\url{http://gigantum.com/}) is a platform for open
  and decentralized data science with a focus on using automation and
  user-friendly tools for easy sharing of reproducible computational
  workflows. Gigantum builds on the \emph{Gigantum Client} (running
  either locally or on a remote server) for development and execution of
  data-focused \emph{Projects}, which can be stored and shared via the
  \emph{Gigantum Hub} or via a zipfile export. The Client is a
  user-friendly interface to a backend using Docker containers to
  package, build, and run Gigantum projects. It is configured to use a
  default set of Docker base images
  (\url{https://github.com/gigantum/base-images}), and users are able to
  define and configure their own custom images. The available images
  include two with R based on Ubuntu Linux and these have the
  \href{https://launchpad.net/~marutter/+archive/ubuntu/c2d4u3.5/}{\texttt{c2d4u}
  CRAN PPA} pre-configured for installation of binary R
  packages\footnote{\href{https://docs.gigantum.com/docs/using-r}{https://docs.gigantum.com/docs/using-r}}.
  The R images vary in the included authoring environment, i.e., Jupyter
  in \texttt{r-tidyverse} or both Jupyter \& RStudio in
  \texttt{rstudio-server}. The independent image stack can be traced
  back to the Gigantum environment and its features. The R images are
  based on Gigantum's \texttt{python3-minimal} image, originally to keep
  the existing front-end configuration, but also to provide consistent
  Python-to-R interoperability. The \texttt{Dockerfile}s also use build
  args to specify bases, for example for different versions of NVIDIA
  CUDA for GPU
  processing\footnote{See \href{https://github.com/gigantum/base-images/blob/master/_templates/python3-minimal-template/Dockerfile}{https://github.com/gigantum/base-images/blob/master/\_templates/python3-minimal-template/Dockerfile} for the \texttt{Dockerfile} of \texttt{python3-minimal}.},
  so that appropriate GPU drivers can be enabled automatically when
  supported. Furthermore, Gigantum's focus lies on environment
  management via GUI and ensuring a smooth user interaction, e.g., with
  reliable and easy conflict detection and resolution. For this reason,
  project repositories store authoritative package information in a
  separate file per package, allowing Git to directly detect conflicts
  and changes. A Dockerfile is generated from this description that
  inherits from the specified base image, and additional custom Docker
  instructions may be appended by users, though Gigantum's default base
  images do not currently include the \texttt{littler} tool, which is
  used by Rocker to install packages within \texttt{Dockerfile}s.
  Because of these specifics, instructions from \texttt{rocker/r-ubuntu}
  could \emph{not} be readily re-used in this image stack (see
  Section~\nameref{conclusions}). Both approaches enable the
  \texttt{apt} package manager \citep{wikipedia_contributors_apt_2020}
  as an installation method, and this is exposed via the GUI-based
  environment
  management\footnote{See \href{https://docs.gigantum.com/docs/environment-management}{https://docs.gigantum.com/docs/environment-management}}.
  The image build and publication process is scripted with Python and
  JSON template configuration files, unlike Rocker images which rely on
  plain \texttt{Dockerfile}s. A further reason in the creation of an
  independent image stack were project constraints requiring a
  Rocker-incompatible licensing of the \texttt{Dockerfile}s, i.e., the
  MIT License.
\end{itemize}

\hypertarget{capture-and-create-environments}{%
\subsection{Capture and create
environments}\label{capture-and-create-environments}}

\label{envs}

Community-maintained images provide a solid basis so users can meet
their own individual requirements. Several second-order R packages
attempt to streamline the process of creating Docker images and using
containers for specific tasks, such as running tests or rendering
reproducible reports. While authoring and managing an environment with
Docker by hand is possible and feasible for
experts\footnote{See, e.g., this tutorial by RStudio on how to manage environments and package versions and to ensure deterministic image builds with Docker: \href{https://environments.rstudio.com/docker}{https://environments.rstudio.com/docker}.},
the following examples show that when environments become too cumbersome
to create manually, automation is a powerful tool. In particular, the
practice of \emph{version pinning}, with system package managers for
different operating systems and with packages \pkg{remotes} and
\pkg{versions} or by using MRAN for R, can greatly increase the
reproducibility of built images and are common approaches.

\textbf{\CRANpkg{dockerfiler}} is an R package designed for building
\texttt{Dockerfile}s straight from R \citep{cran_dockerfiler}. A
scripted creation of a \texttt{Dockerfile} enables iteration and
automation, for example for packaging applications for deployment (see
\nameref{deployment}). Developers can retrieve system requirements and
package dependencies to write a \texttt{Dockerfile}, for example, by
leveraging the tools available in R to parse a \texttt{DESCRIPTION}
file.

\textbf{\pkg{containerit}}
(\url{https://github.com/o2r-project/containerit/}) attempts to take
this one step further and includes these tools to automatically create a
\texttt{Dockerfile} that can execute a given workflow
\citep{nust_containerit_2019}. \pkg{containerit} accepts an R object of
classes \code{"sessionInfo"} or \code{"session\_info"} as input and
provides helper functions to derive these from workflows, e.g., an R
script or R Markdown document, by analysing the session state at the end
of the workflow. It relies on the \pkg{sysreqs}
(\url{https://github.com/r-hub/sysreqs/}) package and it's mapping of
package system dependencies to platform-specific installation package
names\footnote{See \href{https://sysreqs.r-hub.io/}{https://sysreqs.r-hub.io/}.}.
\pkg{containerit} uses \pkg{stevedore} to streamline the user
interaction and improve the created \texttt{Dockerfile}s, e.g., by
running a container for the desired base image to extract the already
available R packages.

\textbf{\CRANpkg{dockr}} is a similar package focusing on the generation
of Docker images for R packages, in which the package itself and all of
the R dependencies, including local non-CRAN packages, are available
\citep{cran_dockr,kjeldgaard_dockr_2019}. \pkg{dockr} facilitates the
organisation of code in the R package structure and the resulting Docker
image mirrors the package versions of the current R session. Users can
manually add statements for non-R dependencies to the
\texttt{Dockerfile}.

\textbf{\CRANpkg{liftr}} \citep{liftr2019} aims to solve the problem of
persistent reproducible reporting in statistical computing based on the
R Markdown format \citep{xie2018}. The irreproducibility of authoring
environments can become an issue for collaborative documents and
large-scale platforms for processing documents. \pkg{liftr} makes the
dynamic R Markdown document the main and sole workflow control file and
the only file that needs to be shared between collaborators for
consistent environments, e.g.~demonstrated in the DockFlow project
(\url{https://dockflow.org}). It introduces new fields to the document
header, allowing users to manually declare the versioned dependencies
required for rendering the document. The package then generates a
\texttt{Dockerfile} from this metadata and provides a utility function
to render the document inside a Docker container, i.e.,
\code{render\_docker("foo.Rmd")}. An RStudio addin even allows
compilation of documents with the single push of a button.

System dependencies are the domain of Docker, but for a full description
of the computing environment, one must also manage the R version and the
R packages. R versions are available via the versioned Rocker image
stack \citep{RJ-2017-065}.
\href{https://github.com/ColinFay/ronline}{r-online} leverages these
images and provides an app for helping users to detect breaking changes
between different R versions and for historic exploration of R. With a
standalone NodeJS app or
\href{https://srv.colinfay.me/r-online}{r-online}, the user can compare
a piece of code run in two separate versions of R. Internally, r-online
opens one or two Docker instances with the given version of R based on
Rocker images, executes a given piece of code, and returns the result to
the user. Regarding R package management, this can be achieved with
MRAN, or with packages such as \CRANpkg{checkpoint}
\citep{cran_checkpoint} and \CRANpkg{renv} \citep{renv2019}, which can
naturally be applied within images and containers. For example,
\pkg{renv} helps users to manage the state of the R library in a
reproducible way, further providing isolation and portability. While
\pkg{renv} does not cover system dependencies, the \pkg{renv}-based
environment can be transferred into a container either by restoring the
environment based on the main configuration file \texttt{renv.lock} or
by storing the \pkg{renv}-cache on the host and not in the container
\citep{ushey_using_2019}. With both the system dependencies and R
packages consciously managed in a Docker image, users can start using
containers as the \emph{only} environment for their workflows, which
allows them to work independently of physical
computers\footnote{Allowing them to be digital "nomads", cf. J.~Bryan's \href{https://github.com/jennybc/docker-why}{https://github.com/jennybc/docker-why}.}
and to assert a specific degree of confidence in the stability of a
developed software
\citep[cf. \texttt{README.Rmd} in][]{marwick_research_2017}.

\hypertarget{development-debugging-and-testing}{%
\subsection{Development, debugging, and
testing}\label{development-debugging-and-testing}}

\label{development}

Containers can also serve as playgrounds and provide specific or ad hoc
environments for the purposes of developing R packages. These
environments may have specific versions of R, of R extension packages,
and of system libraries used by R extension packages, and all of the
above in a specific combination.

First, such containers can greatly facilitate \textbf{fixing bugs and
code evaluation}, because developers and users can readily start a
container to investigate a bug report or try out a piece of software
\citep[cf.][]{ooms_opencpu_2017}. The container can later be discarded
and does not affect their regular system. Using the Rocker images with
RStudio, these disposable environments lack no development comfort
(cf.~Section~\nameref{compendia}). \citet{ooms_opencpu_2017} describes
how \texttt{docker\ exec} can be used to get a root shell in a container
for customisation during software evaluation without writing a
\texttt{Dockerfile}. \citet{eddelbuettel_debugging_2019} describes an
example of how a Docker container was used to debug an issue with a
package only occurring with a particular version of Fortran, and using
tools which are not readily available on all platforms (e.g., not on
macOS).

Second, the strong integration of \textbf{system libraries in core
packages} in the \href{https://www.r-spatial.org/}{R-spatial community}
makes containers essential for stable and proactive development of
common classes for geospatial data modelling and analysis. For example,
GDAL \citep{gdal_2019} is a crucial library in the geospatial domain.
GDAL is a system dependency allowing R packages such as \CRANpkg{sf},
which provides the core data model for geospatial vector data, or
\CRANpkg{rgdal}, to accommodate users to be able to read and write
hundreds of different spatial raster and vector formats
\citep{pebesma_simple_2018,cran_rgdal}. \pkg{sf} and \pkg{rgdal} have
hundreds of indirect reverse imports and dependencies and, therefore,
the maintainers spend a lot of effort trying not to break them.
Purpose-built Docker images are used to prepare for upcoming releases of
system libraries, individual bug reports, and for the lowest supported
versions of system
libraries\footnote{Cf. \href{https://github.com/r-spatial/sf/tree/master/inst/docker}{https://github.com/r-spatial/sf/tree/master/inst/docker}, \href{https://github.com/Nowosad/rspatial_proj6}{https://github.com/Nowosad/rspatial\_proj6}, and \href{https://github.com/r-spatial/sf/issues/1231}{https://github.com/r-spatial/sf/issues/1231}}.

Third, special-purpose images exist for identifying problems beyond the
mere R code, such as \textbf{debugging R memory problems}. These images
significantly reduce the barriers to following complex steps for fixing
memory allocation bugs \citep[cf. Section~4.3 in][]{core_writing_1999}.
These problems are hard to debug and critical, because when they do
occur they lead to fatal crashes.
\href{https://github.com/rocker-org/r-devel-san}{\texttt{rocker/r-devel-san}}
and
\href{https://github.com/rocker-org/r-devel-san-clang}{\texttt{rocker/r-devel-ubsan-clang}}
are Docker images that have a particularly configured version of R to
trace such problems with gcc and clang compilers, respectively
\citep[cf.~\CRANpkg{sanitizers} for examples,][]{eddelbuettel_sanitizers_2014}.
\href{https://github.com/wch/r-debug}{\texttt{wch/r-debug}} is a
purpose-built Docker image with \emph{multiple} instrumented builds of
R, each with a different diagnostic utility activated.

Fourth, containers are useful for \textbf{testing} R code during
development. To submit a package to CRAN, an R package must work with
the development version of R, which must be compiled locally; this can
be a challenge for some users. The \textbf{R-hub} project provides
\emph{``a collection of services to help R package development''}, with
the package builder as the most prominent one \citep{r-hub_docs_2019}.
R-hub makes it easy to ensure that no errors occur, but fixing errors
still often warrants a local setup, e.g., using the image
\texttt{rocker/r-devel}, as is testing packages with native code, which
can make the process more complex \citep[cf.][]{eckert_building_2018}.
The R-hub Docker images can also be used to debug problems locally using
various combinations of Linux platforms, R versions, and
compilers\footnote{See \href{https://r-hub.github.io/rhub/articles/local-debugging.html}{https://r-hub.github.io/rhub/articles/local-debugging.html} and \href{https://blog.r-hub.io/2019/04/25/r-devel-linux-x86-64-debian-clang/}{https://blog.r-hub.io/2019/04/25/r-devel-linux-x86-64-debian-clang/}}.
The images go beyond the configurations, or \emph{flavours}, used by
CRAN for checking
packages\footnote{\href{https://cran.r-project.org/web/checks/check_flavors.html}{https://cran.r-project.org/web/checks/check\_flavors.html}},
e.g., with CentOS-based images, but they lack a container for checking
on Windows or OS X. The images greatly support package developers to
provide support on operating systems with which they are not familiar.
The package \pkg{dockertest}
(\url{https://github.com/traitecoevo/dockertest/}) is a proof of concept
for automatically generating \texttt{Dockerfile}s and building images
specifically to run
tests\footnote{\pkg{dockertest} is not actively maintained, but mentioned still because of its interesting approach.}.
These images are accompanied by a special launch script so the tested
source code is not stored in the image; instead, the currently checked
in version from a local Git repository is cloned into the container at
runtime. This approach separates the test environment, test code, and
current working copy of the code. \label{rselenium} Another use case
where a container can help to standardise tests across operating systems
is detailed the vignettes of the package \CRANpkg{RSelenium}
\citep{rselenium_2019}. The package recommends Docker for running the
\href{https://selenium.dev/}{Selenium} Server application needed to
execute test suites on browser-based user interfaces and webpages, but
it requires users to manually manage the containers.

\label{ci} Fifth, Docker images can be used \textbf{on continuous
integration (CI) platforms} to streamline the testing of packages.
\citet{ye_docker_2019} describes how they speed up the process of
testing by running tasks on \href{https://travis-ci.org/}{Travis~CI}
within a container using \texttt{docker\ exec}, e.g., the package check
or rendering of documentation. \citet{cardozo_faster_2018} also saved
time with Travis~CI by re-using the testing image as the basis for an
image intended for publication on Docker~Hub.
\href{https://github.com/ColinFay/r-ci}{\texttt{r-ci}} is, in turn, used
with \href{https://docs.gitlab.com/ee/ci/}{GitLab~CI}, which itself is
built on top of Docker images: the user specifies a base Docker image
and control code, and the whole set of tests is run inside a container.
The \texttt{r-ci} image stack combines \texttt{rocker} versioning and a
series of tools specifically designed for testing in a fixed environment
with a customized list of preinstalled packages. Especially for
long-running tests or complex system dependencies, these approaches to
separate installation of build dependencies with code testing streamline
the development process. Containers can also simplify the integration of
R software into larger, multi-language CI pipelines. Furthermore, with
each change, even this manuscript is rendered into a PDF and deployed to
a GitHub-hosted website (see \texttt{.travis.yml} and
\texttt{Dockerfile} in the manuscript repository), not because of
concern about time, but to control the environment used on a CI server.
This gives, on the one hand, easy access after every update of the R
Markdown source code and, on the other hand, a second controlled
environment to make sure that the article renders successfully and
correctly.

\hypertarget{processing}{%
\subsection{Processing}\label{processing}}

The portability of containerised environments becomes particularly
useful for improving expensive processing of data or shipping complex
processing pipelines. First, it is possible to \textbf{offload complex
processing to a server} or clouds and also to execute processes in
parallel to speed up or to serve many users.
\textbf{\CRANpkg{batchtools}} provides a parallel implementation of the
\texttt{Map} function for various schedulers \citep{Lang2017batchtools}.
For example, the package can
\href{https://mllg.github.io/batchtools/reference/makeClusterFunctionsDocker.html}{schedule
jobs with Docker Swarm}. \textbf{\CRANpkg{googleComputeEngineR}} has the
function \code{gce\_vm\_cluster()} to create clusters of 2 or more
virtual machines, running multi-CPU architectures
\citep{googleComputeEngineR_2019}. Instead of running a local R script
with the local CPU and RAM restrictions, the same code can be processed
on all CPU threads of the cluster of machines in the cloud, all running
a Docker container with the same R environments.
\pkg{googleComputeEngineR} integrates with the R parallelisation package
\CRANpkg{future} \citep{cran_future} to enable this with only a few
lines of R
code\footnote{\href{https://cloudyr.github.io/googleComputeEngineR/articles/massive-parallel.html}{https://cloudyr.github.io/googleComputeEngineR/articles/massive-parallel.html}}.
\href{https://cloud.run}{Google Cloud Run} is a CaaS (Containers as a
Service) platform. Users can launch containers using any Docker image
without worrying about underlying infrastructure in a so-called
serverless configuration. The service takes care of network ingress,
scaling machines up and down, authentication, and authorisation---all
features which are non-trivial for a developer to build and maintain on
their own. This can be used to scale up R code to millions of instances
if need be with little or no changes to existing code, as demonstrated
by the proof of concept
\texttt{cloudRunR}\footnote{\href{https://github.com/MarkEdmondson1234/cloudRunR}{https://github.com/MarkEdmondson1234/cloudRunR}},
which uses Cloud Run to create a scalable R-based API using
\CRANpkg{plumber} \citep{cran_plumber}.
\href{https://cloud.google.com/cloud-build/}{Google Cloud Build} and the
Google Container Registry are a continuous integration service and an
image registry, respectively, that offload building of images to the
cloud, while serving the needs of commercial environments such as
private Docker images or image stacks. As Google Cloud Build itself can
run any container, the package \pkg{googleCloudRunner} demonstrates how
R can be used as the control language for one-time or batch processing
jobs and scheduling of
jobs\footnote{\href{https://code.markedmondson.me/googleCloudRunner/articles/cloudbuild.html}{https://code.markedmondson.me/googleCloudRunner/articles/cloudbuild.html}}.
\label{drake} \textbf{\CRANpkg{drake}} is a workflow manager for data
science projects \citep{landau_drake_2019}. It features implicit
parallel computing and automated detection of the parts of the work that
actually needs to be re-executed. \pkg{drake} has been demonstrated to
run inside containers for high
reproducibility\footnote{See for example \href{https://github.com/joelnitta/pleurosoriopsis}{https://github.com/joelnitta/pleurosoriopsis} or \href{https://gitlab.com/ecohealthalliance/drake-gitlab-docker-example}{https://gitlab.com/ecohealthalliance/drake-gitlab-docker-example}, the latter even running in a continuous integration platform (cf.~\nameref{development}.}.
Furthermore, \pkg{drake} workflows have been shown to use \pkg{future}
package's function \code{makeClusterPSOCK()} for sending parts of the
workflow to a Docker image for
execution\footnote{\href{https://docs.ropensci.org/drake/index.html?q=docker\#with-docker}{https://docs.ropensci.org/drake/index.html?q=docker\#with-docker}}
\citep[see package's function documentation;~][]{future_2020}. In the
latter case, the container control code must be written by the user, and
the \pkg{future} package ensures that the host and worker can connect
for communicating over socket connections. \textbf{RStudio Server Pro}
includes a functionality called
\href{https://solutions.rstudio.com/launcher/overview/}{Launcher} (since
version~1.2, released in 2019). It gives users the ability to spawn R
sessions and background/batch jobs in a scalable way on external
clusters, e.g.,
\href{https://support.rstudio.com/hc/en-us/articles/360019253393-Using-Docker-images-with-RStudio-Server-Pro-Launcher-and-Kubernetes}{Kubernetes
based on Docker images} or \href{https://slurm.schedmd.com/}{Slurm}
clusters, and optionally, with Singularity containers. A benefit of the
proprietary Launcher software is the ability for R and Python users to
leverage containerisation's advantages in RStudio without writing
specific deployment scripts or learning about Docker or managing
clusters at all.

\label{pipelines} Second, containers are perfectly suited for
\textbf{packaging and executing software pipelines} and required data.
Containers allow for building complex processing pipelines that are
independent of the host programming language. Due to its original use
case (see~\nameref{introduction}), Docker has no standard mechanisms for
chaining containers together; it lacks definitions and protocols for how
to use environment variables, volume mounts, and/or ports that could
enable the transfer of input (parameters and data) and output (results)
to and from containers. Some packages, e.g., \pkg{containerit}, provide
Docker images that can be used very similar to a CLI, but this usage is
cumbersome\footnote{\href{https://o2r.info/containerit/articles/container.html}{https://o2r.info/containerit/articles/container.html}}.
\textbf{\pkg{outsider}} (\url{https://docs.ropensci.org/outsider/})
tackles the problem of integrating external programs into an R workflow
without the need for users to directly interact with containers
\citep{bennett_outsider_2020}. Installation and usage of external
programs can be difficult, convoluted and even impossible if the
platform is incompatible. Therefore, \pkg{outsider} uses the
platform-independent Docker images to encapsulate processes in
\emph{outsider modules}. Each outsider module has a \texttt{Dockerfile}
and an R package with functions for interacting with the encapsulated
tool. Using only R functions, an end-user can install a module with the
\pkg{outsider} package and then call module code to seamlessly integrate
a tool into their own R-based workflow. The \pkg{outsider} package and
module manage the containers and handle the transmission of arguments
and the transfer of files to and from a container. These functionalities
also allow a user to launch module code on a remote machine via SSH,
expanding the potential computational scale. Outsider modules can be
hosted code-sharing services, e.g., on GitHub, and \pkg{outsider}
contains discovery functions for them.

\hypertarget{deployment-and-continuous-delivery}{%
\subsection{Deployment and continuous
delivery}\label{deployment-and-continuous-delivery}}

\label{deployment}

The cloud is the natural environment for containers, and, therefore,
containers are the go-to mechanism for deploying R server applications.
More and more continuous integration (CI) and continuous delivery (CD)
services also use containers, opening up new options for use. The
controlled nature of containers, i.e., the possibility to abstract
internal software environment from a minimal dependency outside of the
container is crucial, for example to match test or build environments
with production environments or transfer runnable entities to
as-a-service infrastructures.

First, different packages use containers for the \textbf{deployment of R
and \pkg{Shiny} apps}. \pkg{Shiny} is a popular package for creating
interactive online dashboards with R, and it enables users with very
diverse backgrounds to create stable and user-friendly web applications
\citep{cran_shiny}. \emph{ShinyProxy} (\url{https://www.shinyproxy.io/})
is an open-source tool to deploy Shiny apps in an enterprise context,
where it features single sign-on, but it can also be used in scientific
use cases
\citep[e.g., ][]{savini_epiexplorer_2019,glouzon_structurexplor_2017}.
ShinyProxy uses Docker containers to isolate user sessions and to
achieve scalability for multi-user scenarios with multiple apps.
ShinyProxy itself is written in Java to accommodate corporate
requirements and may itself run in a container for stability and
availability. The tool is built on ContainerProxy
(\url{https://www.containerproxy.io/}), which provides similar features
for executing long-running R jobs or interactive R sessions. The started
containers can run on a regular Docker host but also in clusters.
Continuous integration and deployment (CI/CD) for Shiny applications
using Shinyproxy can be achieved, e.g., via GitLab pipelines or with a
combination of GitHub and Docker Hub. A pipeline can include building
and checking R packages and Shiny apps. After the code has passed the
checks, Docker images are built and pushed to the container registry.
The pipeline finishes with triggering a
\href{https://github.com/analythium/shinyproxy-1-click/blob/master/digitalocean/setup.md\#setting-up-webhook}{webhook}
on the server, where the deployment script is executed. The script can
update configurations or pull the new Docker images. There is a
\href{https://marketplace.digitalocean.com/apps/shinyproxy}{ShinyProxy
1-Click App} in the DigitalOcean marketplace that is set up with these
webhooks. The
\href{https://github.com/analythium/shinyproxy-1-click/blob/master/digitalocean/secure.md}{documentation}
explains how to set up HTTPS with ShinyProxy and webhooks.

Another example is the package \CRANpkg{golem}, which makes heavy use of
\pkg{dockerfiler} when it comes to creating the \texttt{Dockerfile} for
building and deploying production-grade Shiny applications
\citep{cran_golem}. \CRANpkg{googleComputeEngineR} enables quick
deployments of key R services, such as RStudio and Shiny, onto cloud
virtual machines (VMs) with Google Cloud Compute Engine
\citep{googleComputeEngineR_2019}. The package utilises
\texttt{Dockerfile}s to move the labour of setting up those services
from the user to a premade Docker image, which is configured and run in
the cloud VM. For example, by specifying the template
\texttt{template="rstudio"} in functions \code{gce\_vm\_template()} and
\code{gce\_vm()} an up-to-date RStudio Server image is launched for
development work, whereas specifying \texttt{template="rstudio-gpu"}
will launch an RStudio Server image with a GPU attached, etc.

Second, containers can be used to create \textbf{platform installation
packages} in a DevOps setting. The
\href{https://www.opencpu.org/}{OpenCPU} system provides an HTTP API for
data analysis based on R. \citet{ooms_opencpu_2017} describes how
various platform-specific installation files for OpenCPU are created
using Docker~Hub. The automated builds install the software stack from
the source code on different operating systems; afterwards a script file
downloads the images and extracts the OpenCPU binaries.

Third, containers can greatly facilitate the \textbf{deployment to
existing infrastructures}. \emph{Kubernetes}
(\url{https://kubernetes.io/}) is a container-orchestration system for
managing container-based application deployment and scaling. A
\emph{cluster} of containers, orchestrated as a single deployment, e.g.,
with Kubernetes, can mitigate limitations on request volumes or a
container occupied with a computationally intensive task. A cluster
features load-balancing, autoscaling of containers across numerous
servers (in the cloud or on premise), and restarting failed ones. Many
organisations already use a Kubernetes cluster for other applications,
or a managed cluster can be acquired from service providers. Docker
containers are used within Kubernetes clusters to hold native code, for
which Kubernetes creates a framework around network connections and
scaling of resources up and down. Kubernetes can thereby host R
applications, big parallel tasks, or scheduled batch jobs in a scalable
way, and the deployment can even be triggered by changes to code
repositories \citep[i.e., CD, see][]{edmondson_r_2018}. The package
\pkg{googleKubernetesR}
(\url{https://github.com/RhysJackson/googleKubernetesR}) is a proof of
concept for wrapping the Google Kubernetes Engine API, Google's hosted
Kubernetes solution, in an easy-to-use R package. The package
\CRANpkg{analogsea} provides a way to programmatically create and
destroy cloud VMs on the \href{https://www.digitalocean.com/}{Digital
Ocean} platform \citep{analogsea_2019}. It also includes R wrapper
functions to install Docker in such a VM, manage images, and control
containers straight from R functions. These functions are translated to
Docker CLI commands and transferred transparently to the respective
remote machine using SSH. \pkg{AzureContainers} is an umbrella package
that provides interfaces for three commercial services of Microsoft's
Azure Cloud, namely
\href{https://azure.microsoft.com/en-us/services/container-instances/}{Container
Instances} for running individual containers,
\href{https://azure.microsoft.com/en-us/services/container-registry/}{Container
Registry} for private image distribution, and
\href{https://azure.microsoft.com/en-us/services/kubernetes-service/}{Kubernetes
Service} for orchestrated deployments. While a package like
\pkg{plumber} provides the infrastructure for turning an R workflow into
a web service, for production purposes it is usually necessary to take
into account scalability, reliability and ease of management.
AzureContainers provides an R-based interface to these features and,
thereby, simplifies complex infrastructure management to a number of R
function calls, given an Azure account with sufficient
credit\footnote{See \emph{"Deploying a prediction service with Plumber"} vignette for details:  \href{https://cran.r-project.org/web/packages/AzureContainers/vignettes/vig01_plumber_deploy.html}{https://cran.r-project.org/web/packages/AzureContainers/vignettes/vig01\_plumber\_deploy.html}.}.
\href{https://www.heroku.com/}{Heroku} is another cloud platform as a
service provider, and it supports container-based applications.
\texttt{heroku-docker-r}
(\url{https://github.com/virtualstaticvoid/heroku-docker-r}) is an
independent project providing a template for deploying R applications
based on Heroku's image stack, including multiple examples for
interfacing R with other programming languages. Yet the approach
requires manual management of the computing environment.

Independent integrations of R for different cloud providers lead to
repeated efforts and code fragmentation. To mitigate these problems and
to avoid vendor lock-in motivated the
\href{https://www.openfaas.com/}{OpenFaaS project}. OpenFaas facilitates
the deployment of functions and microservices to Kubernetes or Docker
Swarm. It is language-agnostic and provides auto-scaling, metrics, and
an API gateway. Reduced boilerplate code is achieved via templates.
Templates for
R\footnote{See OpenFaaS R templates at \href{https://github.com/analythium/openfaas-rstats-templates}{https://github.com/analythium/openfaas-rstats-templates}.}
are provided based on Rocker's Debian and R-hub's
\href{https://github.com/r-hub/r-minimal}{r-minimal} Alpine images. The
templates use
\href{https://docs.docker.com/develop/develop-images/multistage-build/}{multi-stage
Docker builds} to combine R base images with the OpenFaaS `watchdog', a
tiny Golang web server. The watchdog marshals an HTTP request and
invokes the actual application. The R session uses \pkg{plumber} or
similar packages for the API endpoint with packages and data preloaded,
thus minimizing response times.

The prevalence of Docker in industry naturally leads to the use of R in
containers, as companies already manage platforms in Docker containers.
These products often entail a large amount of open-source software in
combination with proprietary layers adding the relevant
commercialisation features. One such example is RStudio's data science
platform \href{https://rstudio.com/products/team/}{RStudio Team}. It
allows teams of data scientists and their respective IT/DevOps groups to
develop and deploy code in R and Python around the RStudio Open-Source
Server inside of Docker images, without requiring users to learn new
tools or directly interact with containers. The best practices for
\href{https://support.rstudio.com/hc/en-us/articles/360021594513-Running-RStudio-with-Docker-containers}{running
RStudio with Docker containers} as well as
\href{https://github.com/rstudio/rstudio-docker-products}{Docker images}
for RStudio's commercial products are publicly available.

\hypertarget{using-r-to-power-enterprise-software-in-production-environments}{%
\subsection{Using R to power enterprise software in production
environments}\label{using-r-to-power-enterprise-software-in-production-environments}}

\label{enterprise}

R has been historically viewed as a tool for analysis and scientific
research, but not for creating software that corporations can rely on
for production services. However, thanks to advancements in R running as
a web service, along with the ability to deploy R in Docker containers,
modern enterprises are now capable of having real-time machine learning
powered by R. A number of packages and projects have enabled R to
respond to client requests over TCP/IP and local socket servers, such as
\CRANpkg{Rserve} \citep{cran_rserve}, \CRANpkg{svSocket}
\citep{grosjean_sciviews_2019}, \href{http://www.rapache.net}{rApache}
and more recently \pkg{plumber} (\url{https://www.rplumber.io/}) and
\pkg{RestRserve} (\url{http://restrserve.org}), which even processes
incoming requests in parallel with forked processes using
\CRANpkg{Rserve}. The latter two also provide documentation for
deployment with Docker or ready-to-use images with automated
builds\footnote{See \href{https://www.rplumber.io/docs/hosting.html\#docker}{https://www.rplumber.io/docs/hosting.html\#docker}, \href{https://hub.docker.com/r/trestletech/plumber/}{https://hub.docker.com/r/trestletech/plumber/} and \href{https://hub.docker.com/r/rexyai/restrserve/}{https://hub.docker.com/r/rexyai/restrserve/}.}.
These software allow other (remote) processes and programming languages
to interact with R and to expose R-based function in a service
architecture with HTTP APIs. APIs based on these packages can be
deployed with scalability and high availability using containers. This
pattern of deploying code matches those used by software engineering
services created in more established languages in the enterprise domain,
such as Java or Python, and R can be used alongside those languages as a
first-class member of a software engineering technical stack.

CARD.com implemented a web application for the optimisation of the
acquisition flow and the real-time analysis of debit card transactions.
The software used \pkg{Rserve} and rApache and was deployed in Docker
containers. The R session behind \pkg{Rserve} acted as a read-only
in-memory database, which was extremely fast and scalable, for the many
concurrent rApache processes responding to the live-scoring requests of
various divisions of the company. Similarly deodorised R scripts were
responsible for the ETL processes and even the client-facing email, text
message and push notification alerts sent in real-time based on card
transactions. The related Docker images were made available at
\url{https://github.com/cardcorp/card-rocker}. The images extended
\texttt{rocker/r-base} and additionally entailed an SSH client and a
workaround for being able to mount SSH keys from the host, Pandoc, the
Amazon Web Services (AWS) SDK, and Java, which is required by the AWS
SDK. The AWS SDK allowed for running R consumers reading from real-time
data processing streams of \href{https://aws.amazon.com/kinesis/}{AWS
Kinesis}
\footnote{See useR!2017 talk \href{https://static.sched.com/hosted\_files/user2017/2f/AWR Kinesis at useR 2017.pdf}{"Stream processing with R in AWS"}.}.
The applications were deployed on Amazon Elastic Container Service
(\href{https://aws.amazon.com/ecs/}{ECS}). The main takeaways from using
R in Docker were not only that pinning the R package versions via MRAN
is important, but also that moving away from Debian testing to a
distribution with long-term support can be necessary. For the use case
at hand, this switch allowed for more control over upstream updates and
for minimising the risk of breaking the automated builds of the Docker
images and production jobs.

The AI @ T-Mobile team created a set of machine learning models for
natural language processing to help customer care agents manage
text-based messages from customers \citep{t-mobile_enterprise_2018}. For
example, one model identifies whether a message is from a customer
\citep[see \pkg{Shiny}-based \href{https://secure.message.t-mobile.com/v1/shiny/is-customer/app/}{demo} further described by ][]{nolis_small_2019},
and others tell which customers are likely to make a repeat purchase. If
a data scientist creates a such a model and exposes it through a
\pkg{plumber} API, then someone else on the marketing team can write
software that sends different emails depending on that real-time
prediction. The models are convolutional neural networks that use the
\CRANpkg{keras} package \citep{cran_keras} and run in a Rocker
container. The corresponding \texttt{Dockerfile}s are published
\href{https://github.com/tmobile/r-tensorflow-api}{on GitHub}. Since the
models power tools for agents and customers, they need to have extremely
high uptime and reliability. The AI @ T-Mobile team found that the
models performed well, and today these models power real-time services
that are called over a million times a day.

\hypertarget{common-or-public-work-environments}{%
\subsection{Common or public work
environments}\label{common-or-public-work-environments}}

\label{workenvs}

The fact that Docker images are portable and well defined make them
useful when more than one person needs access to the same computing
environment. This is even more useful when some of the users do not have
the expertise to create such an environment themselves, and when these
environments can be run in public or using shared infrastructure. For
example, \emph{RCloud} (\url{https://rcloud.social}) is a cloud-based
platform for data analysis, visualisation and collaboration using R. It
provides a \texttt{rocker/drd} base image for easy evaluation of the
platform\footnote{\href{https://github.com/att/rcloud/tree/master/docker}{https://github.com/att/rcloud/tree/master/docker}}.

\label{binder} The
\href{https://mybinder.readthedocs.io/en/latest/}{\textbf{Binder}}
project, maintained by the team behind Jupyter, makes it possible for
users to \textbf{create and share computing environments} with others
\citep{jupyter_binder_2018}. A \emph{BinderHub} allows anyone with
access to a web browser and an internet connection to launch a temporary
instance of these custom environments and execute any workflows
contained within. From a reproducibility standpoint, Binder makes it
exceedingly easy to compile a paper, visualize data, and run small
examples from papers or tutorials without the need for any local
installation. To set up Binder for a project, a user typically starts at
an instance of a BinderHub and passes the location of a repository with
a workspace, e.g., a hosted Git repository, or a data repository like
Zenodo. Binder's core internal tool is \texttt{repo2docker}. It
deterministically builds a Docker image by parsing the contents of a
repository, e.g., project dependency configurations or simple
configuration
files\footnote{See supported file types at \href{https://repo2docker.readthedocs.io/en/latest/config\_files.html}{https://repo2docker.readthedocs.io/en/latest/config\_files.html}. For R, the }.
In the most powerful case, \texttt{repo2docker} builds a given
\texttt{Dockerfile}. While this approach works well for most
run-of-the-mill Python projects, it is not so seamless for R projects.
This is partly because \texttt{repo2docker} does not support arbitrary
base images due to the complex auto-generation of the
\texttt{Dockerfile} instructions.

Two approaches make using Binder easier for R users. First,
\textbf{\pkg{holepunch}} (\url{https://github.com/karthik/holepunch}) is
an R package that was designed to make sharing work environments
accessible to novice R users based on Binder. For any R projects that
use the Tidyverse suite \citep{wickham_welcome_2019}, the time and
resources required to build all dependencies from source can often time
out before completion, making it frustrating for the average R user.
\pkg{holepunch} removes some of these limitations by leveraging Rocker
images that contain the Tidyverse along with special Jupyter
dependencies, and only installs additional packages from CRAN and
Bioconductor that are not already part of these images. It short
circuits the configuration file parsing in \texttt{repo2docker} and
starts with the Binder/Tidyverse base images, which eliminates a large
part of the build time and, in most cases, results in a Binder instance
launching within a minute. \pkg{holepunch} also creates a
\texttt{DESCRIPTION} file for essential metadata and dependency
specification, and thereby turns any project into a research compendium
(see~\nameref{compendia}). The \texttt{Dockerfile} included with the
project can also be used to launch an RStudio Server instance locally,
i.e., independent of Binder, which is especially useful when more or
special computational resources can be provided there. The local image
usage reduces the number of separately managed environments and,
thereby, reduces work and increases portability and reproducibility.

Second, the \textbf{Whole~Tale} project (\url{https://wholetale.org})
combines the strengths of the Rocker Project's curated Docker images
with \texttt{repo2docker}. Whole~Tale is a National Science Foundation
(NSF) funded project developing a scalable, open-source, multi-user
platform for reproducible research \citep{brinckman2019, chard2019a}. A
central goal of the platform is to enable researchers to easily create
and publish executable research
objects\footnote{In Whole~Tale a \emph{tale} is a research object that contains metadata, data (by copy or reference), code, narrative, documentation, provenance, and information about the computational environment to support computational reproducibility.}
associated with published research \citep{chard2019b}. Using Whole~Tale,
researchers can create and publish Rocker-based reproducible research
objects to a growing number of repositories including DataONE member
nodes, Zenodo and soon Dataverse. Additionally, Whole~Tale supports
automatic data citation and is working on capabilities for image
preservation and provenance capture to improve the transparency of
published computational research artefacts
\citep{mecum2018, mcphillips2019}. For R users, Whole~Tale extends the
Jupyter Project's \texttt{repo2docker} tool to simplify the
customisation of R-based environments for researchers with limited
experience with either Docker or Git. Multiple options have been
discussed to allow users to change the Ubuntu LTS (long-term support,
currently \emph{Bionic Beaver}) base image,
\texttt{buildpack-deps:bionic}, used in \texttt{repo2docker}. Whole~Tale
implemented a custom
\texttt{RockerBuildPack}\footnote{See \href{https://github.com/whole-tale/repo2docker\_wholetale}{https://github.com/whole-tale/repo2docker\_wholetale}.}.
The build pack combines a \texttt{rocker/geospatial} image with
\texttt{repo2docker}'s
composability\footnote{Composability refers to the ability to combine multiple package managers and their configuration files, such as R, `pip`, and `conda`; see Section~\nameref{binder} for details.}.
This works because both Rocker images and the \texttt{repo2docker} base
image use distributions with APT \citep{wikipedia_contributors_apt_2020}
so that the instructions created by the latter work because of the
compatible shell and package manager.

In \textbf{high-performance computing}, one use for containers is to run
workflows on shared local hardware where teams manage their own
high-performance servers. This can follow one of several design
patterns: Users may deploy containers to hardware as a work environment
for a specific project, containers may provide per-user persistent
environments, or a single container can act as a common multi-user
environment for a server. In all cases, though, the containerised
approach provides several advantages: First, users may use the same
image and thus work environment on desktop and laptop computers. The
first to patterns provide modularity, while the last approach is most
similar to a simple shared server. Second, software updates can be
achieved by updating and redeploying the container rather than by
tracking local installs on each server. Third, the containerised
environment can be quickly deployed to other hardware, cloud or local,
if more resources are necessary or in case of server destruction or
failure. In any of these cases, users need a method to interact with the
containers, be it an IDE exposed over an HTTP port or command-line
access via tools such as SSH. A suitable method must be added to the
container recipes. The Rocker~Project provides containers pre-installed
with the RStudio~IDE. In cases where users store nontrivial amounts of
data for their projects, the data needs to persist beyond the life of
the container. This may be in shared disks, attached network volumes, or
in separate storage where it is uploaded between sessions. In the case
of shared disks or network-attached volumes, care must be taken to match
user permissions, and of course backups are still necessary.

\href{https://cyverse.org}{CyVerse} is an open-source, NSF-funded
cyberinfrastructure platform for the life sciences providing easy access
to computing and storage resources \citep{merchant_iplant_2016}. CyVerse
has a browser-based `data science workbench' called the
\href{https://cyverse.org/discovery-environment}{Discovery Environment}
(DE). The DE uses a combination of
\href{https://research.cs.wisc.edu/htcondor/}{HTCondor} and Kubernetes
for orchestrating container-based analysis and integrates with external
HPC, i.e., \href{https://www.xsede.org/}{NSF-XSEDE}, through
\href{https://www.tacc.utexas.edu/tapis}{TAPIS} (TACC-API's). CyVerse
hosts a multi-petabyte Data Store based on
\href{https://irods.org/}{iRODS} with shared access by its users. The DE
runs Docker containers on demand, with users able to integrate bespoke
containers from DockerHub or other registries
\citep{devisetty_bringing_2016}. Rocker image integration in the DE is
designed to provide researchers with scalable, compute-intensive, R
analysis capabilities for large and complex datasets (e.g.,
genomics/multi-omics, GWAS, phenotypic data, geospatial data, etc.).
These capabilities give users flexibility similar to Binder, but allow
containers to be run on larger computational resources (RAM, CPU, Disk,
GPU), and for longer periods of time (days to weeks). The Rocker
Project's RStudio and Shiny are integrated into the DE by deriving new
images from Rocker
images\footnote{See \href{https://github.com/cyverse-vice/}{https://github.com/cyverse-vice/} for \texttt{Dockerfile}s and configuration scripts; images are auto-built on DockerHub at \href{https://hub.docker.com/u/cyversevice}{https://hub.docker.com/u/cyversevice}.}.
These new images include a reverse proxy using \texttt{nginx} to handle
communication with CyVerse's authentication system
\citep{rstudio_proxy_2020}; CyVerse also allows owners to invite other
registered users to securely access the same instance. The CyVerse
Rocker images further include tools for connecting to its Data Store,
such as the CLI utility \texttt{icommands} for iRODS. CyVerse accounts
are free (with some limitations for non-US users), and the
\href{https://learning.cyverse.org/}{CyVerse Learning Center} provides
community members with information about the platform, including
training and education opportunities.

\label{rocker-gpu} Using \textbf{GPUs} (graphical processing units) as
specialised hardware from containerised common work environments is also
possible and useful \citep{haydel_enhancing_2015}. GPUs are increasingly
popular for compute-intensive machine learning (ML) tasks, e.g., deep
artificial neural networks \citep{schmidhuber_deep_2015}. Although in
this case containers are not completely portable between hardware
environments, but the software stack for ML with GPUs is so complex to
set up that a ready-to-use container is helpful. Containers running GPU
software require drivers and libraries specific to GPU models and
versions, and containers require a specialized runtime to connect to the
underlying GPU hardware. For NVIDIA GPUs, the
\href{https://github.com/NVIDIA/nvidia-docker}{NVIDIA Container Toolkit}
includes a specialized runtime plugin for Docker and a set of base
images with appropriate drivers and libraries. The Rocker~Project
\href{https://github.com/rocker-org/ml}{has a repository} with (beta)
images based on these that include GPU-enabled versions of
machine-learning R packages, e.g., \texttt{rocker/ml} and
\texttt{rocker/tensorflow-gpu}.

\hypertarget{teaching}{%
\subsection{Teaching}\label{teaching}}

Two use cases demonstrate the practical usefulness and advantages of
containerisation in the context of teaching. On the one hand a special
case of shared computing environments (see Section~\ref{workenvs}), and
on the other hand leveraging sandboxing and controlled environments for
auto-grading.

\textbf{Prepared environments for teaching} are especially helpful for
(a) introductory courses, where students often struggle with the first
step of installation and configuration
\citep{cetinkaya-rundel_infrastructure_2018}, and (b) courses that
require access to a relatively complex setup of software tools, e.g.,
database systems. \citet{cetinkaya-rundel_infrastructure_2018} describe
how a Docker-based deployment of RStudio (i) avoided problems with
troubleshooting individual students' computers and greatly increased
engagement through very quickly showing tangible outcomes, e.g., a
visualisation, and (ii) reduced demand on teaching and IT staff. Each
student received access to a personal RStudio instance running in a
container after authentication with the university login, which gives
the benefits of sandboxing and the possibility of limiting resources.
\citet{cetinkaya-rundel_infrastructure_2018} found that for the courses
at hand, actual usage of the UI is intermittent so a single cloud-based
VM with four cores and 28~GB RAM sufficed for over 100 containers. An
example for mitigating \emph{complex setups} is teaching databases. R is
very useful tool for interfacing with databases, because almost every
open-source and proprietary database system has an R package that allows
users to connect and interact with it. This flexibility is even
broadened by \CRANpkg{DBI} \citep{cran_dbi}, which allows for creating a
common API for interfacing these databases, or the \CRANpkg{dbplyr}
package \citep{cran_dbplyr}, which runs \CRANpkg{dplyr}
\citep{cran_dplyr} code straight against the database as queries. But
learning and teaching these tools comes with the cost of deploying or
having access to an environment with the software and drivers installed.
For people teaching R, it can become a barrier if they need to install
local versions of database drivers or connect to remote instances which
might or might not be made available by IT services. Giving access to a
sandbox for the most common environments for teaching databases is the
idea behind \href{https://github.com/ColinFay/r-db}{\texttt{r-db}}, a
Docker image that contains everything needed to connect to a database
from R. Notably, with \texttt{r-db}, users do not have to install
complex drivers or configure their machine in a specific way. The
\texttt{rocker/tidyverse} base image ensures that users can also readily
use packages for analysis, display, and reporting.

The idea of a common environment and partitioning allows for using
\textbf{containers in teaching for secure execution and automated
testing} of submissions by students. First,
\href{https://dodona.ugent.be}{Dodona} is a web platform developed at
Ghent University that is used to teach students basic programming
skills, and it uses Docker containers to test submissions by students.
This means that both the code testing the students' submissions and the
submission itself are executed in a predictable environment, avoiding
compatibility issues between the wide variety of configurations used by
students. The containerisation is also used to shield the Dodona servers
from bad or even malicious code: memory, time and I/O~limits are used to
make sure students cannot overload the system. The web application
managing the containers communicates with them by sending configuration
information as a JSON document over standard input. Every Dodona Docker
image shares a \texttt{main.sh} file that passes through this
information to the actual testing framework, while setting up some error
handling. The testing process in the Docker containers sends back the
test results by writing a JSON document to its standard output channel.
In June 2019, R support was added to Dodona using an image derived from
the \texttt{rocker/r-base} image that sets up the \texttt{runner} user
and \texttt{main.sh} file expected by
Dodona\footnote{\href{https://github.com/dodona-edu/docker-images/blob/master/dodona-r.dockerfile}{https://github.com/dodona-edu/docker-images/blob/master/dodona-r.dockerfile}}.
It also installs the packages required for the testing framework and the
exercises so that this does not have to happen every time a student's
submission is evaluated. The actual testing of R exercises is done using
a custom framework loosely based on \CRANpkg{testthat}
\citep{wickham_testthat_2011}. During the development of the testing
framework, it was found that the \pkg{testthat} framework did not
provide enough information to its reporter system to send back all the
fields required by Dodona to render its feedback. Right now, multiple
statistics courses are developing exercises to automate the feedback for
their lab classes.

Second, \href{https://github.com/PrairieLearn}{PrairieLearn} is another
example of a Docker-based teaching and testing platform. PrairieLearn is
being developed at the University of Illinois at Urbana-Champaign
\citep{prairielearn:2018} and has been in extensive use across several
faculties along with initial use on some other campuses. It uses Docker
containers as key components, both internally for its operations
(programmed mainly in Python as well as in Javascript), as well as for
two reference containers providing, respectively, Python and R
auto-graders. A key design decision made by PrairieLearn permits
\emph{external} grading containers to be supplied and accessed via a
well-defined interface of invoking, essentially, a single script,
\texttt{run.sh}. This script relies on a well-defined file layout
containing JSON-based configurations, support files, exam questions,
supplementary data, and student submissions. It returns per-question
evaluations as JSON result files, which PrarieLearn evaluates,
aggregates and records in a database. The
\href{https://stat430.com}{Data Science Programming Methods} course
\citep{stat430:2019} uses this via the custom
\href{https://github.com/stat430dspm/rocker-pl}{\texttt{rocker-pl}}
container
\citep{rocker-pl:2019}.\footnote{The reference R container was unavailable at the time, and also relies on a heavier CentOS-based build so that a lighter alternative was established.}
The \texttt{rocker-pl} image extends \texttt{rocker/r-base} with the
\pkg{plr} R package \citep{pkg:plr:2019} for integration into
PrarieLearn testing and question evaluation, along with the actual R
packages used in instruction and testing for the course in question. As
automated grading of submitted student answers is close to the
well-understood problem of unit testing, the \CRANpkg{tinytest} package
\citep{CRAN:tinytest} is used for both its core features for testing as
well as clean extensibility. The package \CRANpkg{ttdo}
\citep{CRAN:ttdo} utilizes the extensibility of \pkg{tinytest} to
display context-sensitive colourized differences between incorrect
answers and reference answers using the \CRANpkg{diffobj} package
\citep{CRAN:diffobj}. Additionally, \CRANpkg{ttdo} addresses the issue
of insufficient information collection that Dodona faced by allowing for
the collection of arbitrary, test specific attributes for additional
logging and feedback. The setup, described in more detail by
\citet{paper:r_autograder}, is an excellent illustration of both the
versatility and flexibility offered by Docker-based approaches in
teaching and testing.

\hypertarget{packaging-research-reproducibly}{%
\subsection{Packaging research
reproducibly}\label{packaging-research-reproducibly}}

\label{compendia}

Containers provide a high degree of isolation that is often desirable
when attempting to capture a specific computational environment so that
others can reproduce and extend a research result. Many computationally
intensive research projects depend on specific versions of original and
third-party software packages in diverse languages, joined together to
form a pipeline through which data flows. New releases of even just a
single piece of software in this pipeline can break the entire workflow,
making it difficult to find the error and difficult for others to reuse
existing pipelines. These breakages can make the original the results
irreproducible and, and the chance of a substantial disruption like this
is high in a multi-year research project where key pieces of third-party
software may have several major updates over the duration of the
project. The classical ``paper'' article is insufficient to adequately
communicate the knowledge behind such research projects
\citep[cf.][]{donoho_invitation_2010,marwick_how_2015}.

\citet{gentleman_statistical_2007} coined the term \textbf{Research
Compendium} for a dynamic document together with supporting data and
code. They used the R package system \citep{core_writing_1999} for the
functional prototype all the way to structuring, validating, and
distributing research compendia. This concept has been taken up and
extended\footnote{See full literature list at \href{https://research-compendium.science/}{https://research-compendium.science/}.},
not in the least by applying containerisation and other methods for
managing computing environments---see Section~\nameref{envs}. Containers
give the researcher an isolated environment to assemble these research
pipelines with specific versions of software to minimize problems with
breaking changes and make workflows easier to share
\citep[cf.][]{boettiger_introduction_2015,marwick_packaging_2018}.
Research workflows in containers are safe from contamination from other
activities that occur on the researcher's computer, for example the
installation of the newest version of packages for teaching
demonstrations or specific versions for evaluation of others' works.
Given the users in this scenario, i.e., often academics with limited
formal software development training, templates and assistance with
containers around research compendia is essential. In many fields, we
see that a typical unit of research for a container is a research report
or journal article, where the container holds the compendium, or
self-contained set of data (or connections to data elsewhere) and code
files needed to fully reproduce the article
\citep{marwick_packaging_2018}. The package \pkg{rrtools}
(\url{https://github.com/benmarwick/rrtools}) provides a template and
convenience functions to apply good practices for research compendia,
including a starter \texttt{Dockerfile}. Images of compendium containers
can be hosted on services such as Docker~Hub for convenient sharing
among collaborators and others. Similarly, packages such as
\pkg{containerit} and \pkg{dockerfiler} can be used to manage the
\texttt{Dockerfile} to be archived with a compendium on a data
repository (e.g.~\href{https://zenodo.org/}{Zenodo},
\href{https://dataverse.org/}{Dataverse},
\href{https://figshare.com/}{Figshare}, \href{https://osf.io/}{OSF}). A
typical compendium's \texttt{Dockerfile} will pull a rocker image fixed
to a specific version of R, and install R packages from the MRAN
repository to ensure the package versions are tied to a specific date,
rather than the most recent version. A more extreme case is the
\emph{dynverse} project
\citep{saelens_comparisonsinglecelltrajectory_2019}, which packages over
50 computational methods with different environments (R, Python, C++,
etc.) in Docker images, which can be executed from R. \emph{dynverse}
uses a CI platform (see~\nameref{ci}) to build Rocker-derived images,
test them, and, if the tests succeed, publish them on Docker~Hub.

Future researchers can download the compendium from the repository and
run the included \texttt{Dockerfile} to build a new image that recreates
the computational environment used to produce the original research
results. If building the image fails, the human-readable instructions in
a \texttt{Dockerfile} are the starting point for rebuilding the
environment. When combined with CI (see~\nameref{ci}), a research
compendium set-up can enable \emph{continuous analysis} with easier
verification of reproducibility and audits trails
\citep{beaulieu-jones_reproducibility_2017}.

Further safeguarding practices are currently under development or not
part of common practice yet, such as preserving images
\citep{emsley_framework_2018}, storing both images and
\texttt{Dockerfile}s \citep[cf.][]{nust_opening_2017}, or pinning system
libraries beyond the tagged base images, which may be seen as stable or
dynamic depending on the applied time scale
\citep[see discussion on \texttt{debian:testing} base image in][]{RJ-2017-065}.
A recommendation of the recent National Academies' report on
\emph{Reproducibility and Replicability in Science} is that journals
\emph{``consider ways to ensure computational reproducibility for
publications that make claims based on computations''}
\citep{NASEM2019}. In fields such as political science and economics,
journals are increasingly adopting policies that require authors to
publish the code and data required to reproduce computational findings
reported in published manuscripts, subject to independent verification
\citep{Jacoby2017,Vilhuber2019,Alvarez2018,Christian2018,Eubank2016,King1995}.
Problems with the computational environment, installation and
availability of software dependencies are common. R is gaining
popularity in these communities, such as for creating a research
compendium. In a sample of 105 replication packages published by the
\emph{American Journal of Political Science} (AJPS), over 65\% use R.
The NSF-funded Whole Tale project, which was mentioned above, uses the
Rocker Project community images with the goal of improving the
reproducibility of published research artefacts and simplifying the
publication and verification process for both authors and reviewers by
reducing errors and time spent specifying the environment.

\hypertarget{conclusions}{%
\section{Conclusions}\label{conclusions}}

This article is a snapshot of the R corner in a universe of applications
built with a many-faced piece of software, Docker. \texttt{Dockerfile}s
and Docker images are the go-to methods for collaboration between roles
in an organisation, such as developers and IT operators, and between
participants in the communication of knowledge, such as researchers or
students. Docker has become synonymous with applying the concept of
containerisation to solve challenges of reproducible environments, e.g.,
in research and in development \& production, and of scalable
deployments because it can easily move processing between machines,
e.g., locally, a cloud provider's VM, another cloud provider's
Container-as-a-Service. Reproducible environments, scalability \&
efficiency, and portability across infrastructures are the common themes
behind R packages, use cases, and applications in this work.

The projects presented above show the growing number of users,
developers, and real-world applications in the community and the
resulting innovations. But the applications also point to the challenges
of keeping up with a continuously evolving landscape. Some use cases
have considerable overlap, which can be expected as a common language
and understanding of good practices is still taking shape. Also, the
ease with which one can create complex software systems with Docker to
serve one's specific needs, such as an independent Docker image stack,
leads to parallel developments. This ease-of-DIY in combination with the
difficulty of reusing parts from or composing multiple
\texttt{Dockerfile}s is a further reason for \textbf{fragmentation}.
Instructions can be outsourced into distributable scripts and then
copied into the image during build, but that makes \texttt{Dockerfile}s
harder to read. Scripts added to a \texttt{Dockerfile} also add a layer
of complexity and increase the risk of incomplete recipes. Despite the
different image stacks presented here, the pervasiveness of Rocker
images can be traced back to its maintainers and the user community
valuing collaboration and shared starting points over impulses to create
individual solutions. Aside from that, fragmentation may not be a bad
sign but may instead be a reflection of a growing market that is able to
sustain multiple related efforts. With the maturing of core building
blocks, such as the Rocker suite of images, more working systems will be
built, but they may simply work behind the curtains. Docker alone, as a
flexible core technology, is not a feasible level of collaboration and
abstraction. Instead, the use cases and applications observed in this
work provide a more useful division.

Nonetheless, at least on the level of R packages some
\textbf{consolidation} seems in order, e.g., to reduce the number of
packages creating \texttt{Dockerfile}s from R code or controlling the
Docker daemon with R code. It remains to be seen which approach to
control Docker, via the Docker API as \pkg{stevedore} or via system
calls as \pkg{dockyard}/\pkg{docker}/\pkg{dockr}, is more sustainable,
or whether the question will be answered by the endurance of maintainers
and sufficient funding. Similarly, capturing environments and their
serialisation in form of a \texttt{Dockerfile} currently is happening at
different levels of abstraction, and re-use of functionality seems
reasonable, e.g., \pkg{liftr} could generate the environment with
\pkg{containerit}, which in turn may use \pkg{dockerfiler} for low-level
R objects representing a \texttt{Dockerfile} and its instructions. In
this consolidation of R packages, the Rocker~Project could play the role
of a coordinating entity. Nonetheless, for the moment, it seems that the
Rocker Project will focus on maintaining and extending its image stacks,
e.g., images for GPU-based computing and artificial intelligence. Even
with coding being more and more accepted as a required and achievable
skill, an easier access, for example by exposing containerisation
benefits via simple user interfaces in the users' IDE, could be an
important next step, since currently containerisation happens more in
the background for UI-based development (e.g., a \texttt{rocker/rstudio}
image in the cloud). Furthermore, the maturing of the Rockerverse
packages for managing containers may lead to them being adopted in
situations where manual coding is currently required, e.g.~in the case
of \pkg{RSelenium} or \pkg{drake} (see Sections~\nameref{rselenium} and
\nameref{drake} respectively). In some cases, e.g., for \pkg{analogsea},
the interaction with the Docker daemon may remain too specific to re-use
first-order packages to control Docker.

New features which make complex workflows accessible and reproducible
and the variety in packages connected with containerisation, even when
they have overlapping features, are a signal and support for a growing
user base. This growth is possibly the most important goal for the
foreseeable future in the \emph{Rockerverse}, and, just like the Rocker
images have matured over years of use and millions of runs, the new
ideas and prototypes will have to prove themselves. It should be noted
that the dominant position is that Docker is a blessing and a curse for
these goals. It might be wise to start experimenting with non-Docker
containerisation tools now, e.g., R packages interfacing with other
container engines, such as
\href{https://github.com/containers/libpod}{podman/buildah}, or an R
package for creating \texttt{Singularity} files. Such efforts might help
to avoid lock-in and to design sustainable workflows based on concepts
of \emph{containerisation}, not on their implementation in Docker. If
adoption of containerisation and R continue to grow, the missing pieces
for a success predominantly lie in (a) coordination and documentation of
activities to reduce repeated work in favour of open collaboration, (b)
the sharing of lessons learned from use cases to build common knowledge
and language, and (c) a sustainable continuation and funding for
development, community support, and education. A first concrete effort
to work towards these missing pieces should be sustaining the structure
and captured status quo from this work in the form of a \emph{CRAN Task
View on containerisation}.

\hypertarget{author-contributions}{%
\section{Author contributions}\label{author-contributions}}

The ordering of authors following DN and DE is alphabetical. DN
conceived the article idea,
\href{https://github.com/nuest/rockerverse-paper/issues/3}{initialised the formation of the writing team},
wrote sections not mentioned below, and revised all sections. DE wrote
the introduction and the section about containerisation and the Rocker
Project, and reviewed all sections. DB wrote the section on
\pkg{outsider}. GD contributed the CARD.com use case. RC contributed to
the section on interfaces for Docker in R (\emph{dynverse} and
\texttt{dynwrap}). DC contributed content on Gigantum. ME contributed to
the section on processing and deployment to cloud services. CF wrote
paragraphs about \pkg{r-online}, \pkg{dockerfiler}, \pkg{r-ci} and
\pkg{r-db}. EH contributed content on \pkg{dockyard}. LK contributed
content on \pkg{dockr}. SL contributed content on RStudio's usage of
Docker. BM wrote the section on research compendia and made the project
Binder-ready. HN \& JN co-wrote the section on the T-Mobile use case. KR
wrote the section about \pkg{holepunch}. NR wrote paragraphs about
shared work environments and GPUs. LS \& NT wrote the section on
Bioconductor. PS wrote the paragraphs about CI/CD pipelines with
Shinyproxy 1-Click app and OpenFaaS templates. TS \& JW wrote the
section on CyVerse. CvP wrote the section on the usage of Docker
containers in Dodona. CW wrote the sections on Whole Tale and
contributed content about publication reproducibility audits. NX
contributed content on \pkg{liftr}. All authors approved the final
version. This articles was collaboratively written at
\href{https://github.com/nuest/rockerverse-paper/}{https://github.com/nuest/rockerverse-paper/}.
The
\href{https://github.com/nuest/rockerverse-paper/graphs/contributors}{contributors page}
and
\href{https://github.com/nuest/rockerverse-paper/issues/}{discussion issues}
provide details on the respective contributions.

\hypertarget{acknowledgements}{%
\section{Acknowledgements}\label{acknowledgements}}

DN is supported by the project Opening Reproducible Research
(\href{https://www.uni-muenster.de/forschungaz/project/12343}{o2r})
funded by the German Research Foundation (DFG) under project number
\href{https://gepris.dfg.de/gepris/projekt/415851837}{PE~1632/17-1}. The
funders had no role in data collection and analysis, decision to
publish, or preparation of the manuscript. KR was supported in part by a
grant from The Leona M. and Harry B. Helmsley Charitable Trust, award
number 2016PG-BRI004. LS and NT are supported by US NIH / NHGRI awards
U41HG00405 and U24HG010263. CW is supported by the Whole Tale project
(\url{https://wholetale.org}) funded by the US National Science
Foundation (NSF) under award
\href{https://www.nsf.gov/awardsearch/showAward?AWD_ID=1541450}{OAC-1541450}.
NR is supported in part by the Chan-Zuckerberg Initiative Essential Open
Source Software for Science program. We would like to thank Celeste R.
Brennecka from the Scientific Editing Service of the University of
Münster for her editorial support.

\bibliography{rockerverse}

\address{%
Daniel Nüst\\
University of Münster\\
Institute for Geoinformatics\\ Heisenbergstr. 2\\ 48149 Münster,
Germany\\ \orcid{0000-0002-0024-5046}\\
}
\href{mailto:daniel.nuest@uni-muenster.de}{\nolinkurl{daniel.nuest@uni-muenster.de}}

\address{%
Dirk Eddelbuettel\\
University of Illinois at Urbana-Champaign\\
Department of Statistics\\ Illini Hall, 725 S Wright St\\ Champaign, IL
61820, USA\\ \orcid{0000-0001-6419-907X}\\
}
\href{mailto:dirk@eddelbuettel.com}{\nolinkurl{dirk@eddelbuettel.com}}

\address{%
Dom Bennett\\
Gothenburg Global Biodiversity Centre, Sweden\\
Carl Skottsbergs gata 22B\\ 413 19 Göteborg,
Sweden\\ \orcid{0000-0003-2722-1359}\\
}
\href{mailto:dominic.john.bennett@gmail.com}{\nolinkurl{dominic.john.bennett@gmail.com}}

\address{%
Robrecht Cannoodt\\
Ghent University\\
Data Mining and Modelling for Biomedicine group\\ VIB Center for
Inflammation Research\\ Technologiepark 71\\ 9052 Ghent,
Belgium\\ \orcid{0000-0003-3641-729X}\\
}
\href{mailto:robrecht@cannoodt.dev}{\nolinkurl{robrecht@cannoodt.dev}}

\address{%
Dav Clark\\
Gigantum, Inc.\\
1140 3rd Street NE\\ Washington, D.C. 20002,
USA\\ \orcid{0000-0002-3982-4416}\\
}
\href{mailto:dav@gigantum.com}{\nolinkurl{dav@gigantum.com}}

\address{%
Gergely Daróczi\\
\\
\orcid{0000-0003-3149-8537}\\
}
\href{mailto:daroczig@rapporter.net}{\nolinkurl{daroczig@rapporter.net}}

\address{%
Mark Edmondson\\
IIH Nordic A/S, Google Developer Expert for Google Cloud Platform\\
Artillerivej 86\\ 2300 København S,
Denmark\\ \orcid{0000-0002-8434-3881}\\
}
\href{mailto:mark@markedmondson.me}{\nolinkurl{mark@markedmondson.me}}

\address{%
Colin Fay\\
ThinkR\\
5O rue Arthur Rimbaud\\ 93300 Aubervilliers,
France\\ \orcid{0000-0001-7343-1846}\\
}
\href{mailto:contact@colinfay.me}{\nolinkurl{contact@colinfay.me}}

\address{%
Ellis Hughes\\
Fred Hutchinson Cancer Research Center\\
Vaccine and Infectious Disease\\ 1100 Fairview Ave. N., P.O. Box
19024\\ Seattle, WA 98109-1024, USA\\
}
\href{mailto:ehhughes@fredhutch.org}{\nolinkurl{ehhughes@fredhutch.org}}

\address{%
Lars Kjeldgaard\\
Danish Tax Authorities\\
Oestbanegade 123\\ 2100, Koebenhavn Oe\\
}
\href{mailto:lars_kjeldgaard@hotmail.com}{\nolinkurl{lars\_kjeldgaard@hotmail.com}}

\address{%
Sean Lopp\\
RStudio, Inc\\
250 Northern Ave\\ Boston, MA 02210, USA\\
}
\href{mailto:sean@rstudio.com}{\nolinkurl{sean@rstudio.com}}

\address{%
Ben Marwick\\
University of Washington\\
Department of Anthropology\\ Denny Hall 230, Spokane Ln\\ Seattle, WA
98105, USA\\ \orcid{0000-0001-7879-4531}\\
}
\href{mailto:bmarwick@uw.edu}{\nolinkurl{bmarwick@uw.edu}}

\address{%
Heather Nolis\\
T-Mobile\\
12920 Se 38th St.\\ Bellevue, WA, 98006, USA\\
}
\href{mailto:heather.wensler1@t-mobile.com}{\nolinkurl{heather.wensler1@t-mobile.com}}

\address{%
Jacqueline Nolis\\
Nolis, LLC\\
Seattle, WA, USA\\ \orcid{0000-0001-9354-6501}\\
}
\href{mailto:jacqueline@nolisllc.com}{\nolinkurl{jacqueline@nolisllc.com}}

\address{%
Hong Ooi\\
Microsoft\\
Level 5, 4 Freshwater Place\\ Southbank, VIC 3006, Australia\\
}
\href{mailto:hongooi@microsoft.com}{\nolinkurl{hongooi@microsoft.com}}

\address{%
Karthik Ram\\
Berkeley Institute for Data Science\\
University of California\\ Berkeley, CA 94720,
USA\\ \orcid{0000-0002-0233-1757}\\
}
\href{mailto:karthik.ram@berkeley.edu}{\nolinkurl{karthik.ram@berkeley.edu}}

\address{%
Noam Ross\\
EcoHealth Alliance\\
460 W 34th St., Ste. 1701\\ New York, NY 10001,
USA\\ \orcid{0000-0002-2136-0000}\\
}
\href{mailto:ross@ecohealthalliance.org}{\nolinkurl{ross@ecohealthalliance.org}}

\address{%
Lori Shepherd\\
Roswell Park Comprehensive Cancer Center\\
Elm \& Carlton Streets\\ Buffalo, NY, 14263,
USA\\ \orcid{0000-0002-5910-4010}\\
}
\href{mailto:lori.shepherd@roswellpark.org}{\nolinkurl{lori.shepherd@roswellpark.org}}

\address{%
Péter Sólymos\\
Analythium Solutions\\
\#258 150 Chippewa Road\\ Sherwood Park, AB, T8A 6A2,
Canada\\ \orcid{0000-0001-7337-1740}\\
}
\href{mailto:peter@analythium.io}{\nolinkurl{peter@analythium.io}}

\address{%
Tyson Lee Swetnam\\
University of Arizona\\
1657 E Helen St.\\ Tucson, AZ, 85721,
USA\\ \orcid{0000-0002-6639-7181}\\
}
\href{mailto:tswetnam@arizona.edu}{\nolinkurl{tswetnam@arizona.edu}}

\address{%
Nitesh Turaga\\
Roswell Park Comprehensive Cancer Center\\
Elm \& Carlton Streets\\ Buffalo, NY, 14263,
USA\\ \orcid{0000-0002-0224-9817}\\
}
\href{mailto:nitesh.turaga@roswellpark.org}{\nolinkurl{nitesh.turaga@roswellpark.org}}

\address{%
Charlotte Van Petegem\\
Ghent University\\
Department WE02\\ Krijgslaan 281, S9\\ 9000 Gent,
Belgium\\ \orcid{0000-0003-0779-4897}\\
}
\href{mailto:charlotte.vanpetegem@ugent.be}{\nolinkurl{charlotte.vanpetegem@ugent.be}}

\address{%
Jason Williams\\
Cold Spring Harbor Laboratory\\
1 Bungtown Rd.\\ Cold Spring Harbor, NY, 11724,
USA\\ \orcid{0000-0003-3049-2010}\\
}
\href{mailto:williams@cshl.edu}{\nolinkurl{williams@cshl.edu}}

\address{%
Craig Willis\\
University of Illinois at Urbana-Champaign\\
501 E. Daniel St.\\ Champaign, IL 61820,
USA\\ \orcid{0000-0002-6148-7196}\\
}
\href{mailto:willis8@illinois.edu}{\nolinkurl{willis8@illinois.edu}}

\address{%
Nan Xiao\\
Seven Bridges Genomics\\
529 Main St, Suite 6610\\ Charlestown, MA 02129,
USA\\ \orcid{0000-0002-0250-5673}\\
}
\href{mailto:me@nanx.me}{\nolinkurl{me@nanx.me}}

\end{article}

\end{document}